\RequirePackage[2020-02-02]{latexrelease}
\documentclass[preprintnumbers,amsmath,amssymb]{revtex4}
\usepackage{amsmath,amssymb,graphics,epsfig,subfigure}
\usepackage{color}

\newtheorem{prop}{Proposition}
\newtheorem{corollary}{Corollary}

\begin{document}
\renewcommand{\baselinestretch}{1.3}

\title{Partition of free energy for a Brownian quantum oscillator: Effect of dissipation and magnetic field}

\author{Jasleen Kaur  \footnote{jk14@iitbbs.ac.in},
        Aritra Ghosh  \footnote{ag34@iitbbs.ac.in},
        Malay Bandyopadhyay \footnote{malay@iitbbs.ac.in}}

\affiliation{School of Basic Sciences,\\ Indian Institute of Technology Bhubaneswar, Argul, Jatni, Khurda, Odisha 752050, India}

\begin{abstract}
Recently, the quantum counterpart of energy equipartition theorem has drawn considerable attention. Motivated by this, we formulate and investigate an analogous statement for the free energy of a quantum oscillator linearly coupled to a passive heat bath consisting of an infinite number of independent harmonic oscillators. We explicitly demonstrate that the free energy of the Brownian oscillator can be expressed in the form $F(T) = \langle f(\omega,T) \rangle $ where $f(\omega,T)$ is the free energy of an individual bath oscillator. The overall averaging process involves two distinct averages: the first one is over the canonical ensemble for the bath oscillators, whereas the second one signifies averaging over the entire bath spectrum of frequencies from zero to infinity. The latter is performed over a relevant probability distribution function $\mathcal{P}(\omega)$ which can be derived from the knowledge of the generalized susceptibility encountered in linear response theory. The effect of different dissipation mechanisms is also exhibited. We find two remarkable consequences of our results. First, the quantum counterpart of energy equipartition theorem follows naturally from our analysis. The second corollary we obtain is a natural derivation of the third law of thermodynamics for open quantum systems. Finally, we generalize the formalism to three spatial dimensions in the presence of an external magnetic field.
\end{abstract}


\maketitle

\section{Introduction}
The discovery of the fact that thermodynamic principles are consistent with the quantum properties of microscopic systems has led to a flurry of research activity in the recent times. Quantum thermodynamics, as the subject is known (see for example \cite{qt1,qt2} for a pedagogical introduction), has seen several remarkable recent advances such as the formulation of quantum thermodynamic functions \cite{qtf1,qtf2,qtf3,qtf4} and the third law (see \cite{tl1} and references therein), a novel proposal of a quantum analogue of energy equipartition \cite{jarzy0,jarzy1,jarzy2,jarzy3,jarzy31,jarzy4,jarzy5,kaur} as well as studies on non-equilibrium steady states in nanoscale systems \cite{nano1,nano2,nano3}. A particularly interesting aspect is that of the quantum counterpart of the classical energy equipartition theorem, which is being actively worked on. The classical equipartition theorem states that the total kinetic energy $E_k$ of a system is shared equally among all the energetically accessible degrees of freedom at thermal equilibrium with each of them being associated with an energy $\frac{k_BT}{2}$ where $T$ is the absolute temperature. On the other hand, for an open quantum system that interacts with the degrees of freedom of some passive heat bath, the mean energy is found to receive contributions from various bath degrees of freedom which are weighed by an appropriate probability distribution function. Moreover, it has been disclosed that such relevant probability distribution functions are related to the susceptibility tensor from linear response theory which is an experimentally measurable quantity, thus making it viable to verify this novel result from experiments.

\smallskip

In this paper, we consider a prototypical example of an open quantum system, namely, the dissipative oscillator and show that in general, the steady state quantum thermodynamic properties of the system systematically receive contributions from the bath degrees of freedom. Such contributions are dictated by an appropriate probability distribution function \(\mathcal{P}(\omega)\) dependent on the system parameters and the dissipation mechanism under consideration. This feature becomes explicit when one expresses the free energy of the system at temperature \(T\) in the form \(F(T) = \langle f(\omega,T) \rangle\) where \(f(\omega,T)\) is the free energy of an individual bath oscillator at the same temperature. Then it raises the natural question as to whether the function \(\mathcal{P}(\omega)\) is related to the probabilities with which the system receives energy contributions from the bath degrees of freedom in accordance to the recently proposed quantum counterpart of energy equipartition \cite{jarzy0,jarzy1,jarzy2,jarzy3,jarzy31,jarzy4,jarzy5,kaur}. We find that it is indeed the case, i.e. the free energy formulation is consistent with the energy equipartition in all respects with the former generalizing the latter. Further, as a direct consequence of this general formalism, at equilibrium, one can express the entropy of the system as a sum of contributions coming from various bath degrees of freedom in which the third law of thermodynamics emerges in a natural manner. This uncovers a novel feature in the quantum thermodynamics of dissipative systems wherein a probability distribution function \(\mathcal{P}(\omega)\) dependent on the dynamical description of the system or the equations of motion controls how extensive thermodynamic quantities of the system receives contributions from the bath degrees of freedom. Generalization to three dimensions in the presence of an external magnetic field is performed. As examples, we explicitly work out three different dissipation mechanisms, namely Ohmic dissipation, Drude dissipation and radiation bath dissipation.

\smallskip

With this motivation, we present the plan of this paper as follows. In the next section, we consider the Langevin dynamics of a one dimensional quantum particle placed in a harmonic trap with stiffness constant \(K\) and connected to a heat bath through the coordinate variables. The heat bath is taken to be composed of an infinite number of independent harmonic oscillators with varying frequencies obeying a spectral distribution \(J(\omega)\). It is known that the free energy of such a system can be expressed as a difference between the free energy of the interacting system and that of the free bath field (see for example \cite{qtf3}). We assert that this is equivalent to writing \(F(T) = \langle f(\omega,T) \rangle\) where \(f(\omega,T)\) is the free energy of a single bath oscillator and \(\langle \cdot \rangle\) is an averaging performed over all the bath frequencies according to a well defined probability distribution function \(\mathcal{P}(\omega)\). Two important consequences of this are pointed out. In section-(\ref{section3}), we explicitly demonstrate that our assertion is indeed true by considering three different dissipation mechanisms, namely Ohmic, Drude and radiation baths. Section-(\ref{section4}) is about generalization of the formalism to three spatial dimensions together with the possibility of an external magnetic field. It is shown that the assertions made in section-(\ref{section2}) hold even in the three dimensional case. We conclude with some discussions in section-(\ref{section5}).

\section{The model \& formalism}\label{section2}
In this section, we recapitulate some known aspects of the dissipative quantum harmonic oscillator in one dimension and also set up our notation. The system is linearly coupled through coordinate variables with a passive heat bath which is composed of an infinite number of independent quantum harmonic oscillators. Thus, the total Hamiltonian is given by,
\begin{eqnarray}
   H &=& \frac{p^2}{2m} + \frac{1}{2} K x^2 \nonumber \\
   && + \sum_{j}\bigg[\frac{p_j^2}{2m_j} + \frac{1}{2}m_j \omega_j^2 \bigg(q_j - \frac{c_j}{m_j \omega_j^2}x \bigg)^2 \bigg] \label{CLH}
\end{eqnarray}
 where the symbols have their usual meanings. Noting that \(\{x,p\} = i \hbar\) and \(\{q_j,p_k\} = i \hbar \delta_{jk}\), one can derive the Heisenberg equations of motion for the system and bath variables with the system being coupled to the bath. Solving for the bath variables and eliminating them from the equation of motion of the system, one arrives at the following quantum Langevin equation for the dissipative oscillator (see for example \cite{12,12A} and references therein),
\begin{equation}\label{eqnm}
  m \ddot{x}(t) + \int_{-\infty}^{t} \mu(t - t') \dot{x}(t') dt' + K x(t) = F(t)
\end{equation} where,
where \(F(t)\) is an operator valued random force (or quantum noise) and \(\mu(t) = \sum_j m_j \omega_j^2 \cos (\omega_j t) \Theta(t)\) is the friction kernel which by convention is defined to vanish for \(t < 0\) in order to respect the causality principle. The quantum noise is specified by the initial conditions of the bath oscillators making the process non-Markovian. It should perhaps be emphasized that upon if eqn (\ref{CLH}) represents the Hamiltonian of the system and the bath including their interactions, the quantum Langevin equation obtained by eliminating the bath variables from the equation of motion of the system contains an additional term of the form \(m \mu(t) x(0)\). Such a term is known as the initial slip term (IST) and has been absorbed into the definition of the random force as, 
\begin{equation}\label{newnoise}
	F(t) = - m \mu(t) x(0) + K(t)
\end{equation} where \(K(t)\) can be expressed in terms of the initial conditions of the bath variables. Explicitly, it reads, 
\begin{equation}
	K(t) = \sum_{j} c_j\Bigg[ q_j(0) \cos(\omega_jt)+ \frac{p_j(0)}{m_j \omega_j}\sin(\omega_jt)  \Bigg].
\end{equation} This random force operator $K(t)$ is a stationary Gaussian operator noise once the averages of the initial values $q_j(0)$ and $p_j(0)$ are taken with respect to the initial equilibrium density matrix of the bath in canonical ensemble, i.e.
\begin{equation}\label{initialrhouncoupled}
\hat{\rho}_B^{eq}=\frac{1}{Z}\exp\Big\lbrack -\beta \sum_j \Big\lbrace\frac{p_j^{2}(0)}{2m_j}+\frac{1}{2}m_j\omega_j^{2}q_j^{2}(0)\Big\rbrace\Big\rbrack.
\end{equation}
Here, \(Z\) is the canonical partition function of the bath. Now, if we absorb the IST into the quantum noise as shown in eqn (\ref{newnoise}), the shifted noise operator \(F(t)\) depends on the initial position of the system, in addition to the initial conditions of the bath oscillators. Thus, a priori it is not clear whether \(F(t)\) is a Gaussian as is desired. In fact, one finds \(\langle F(t) \rangle \neq 0\) if this average is taken over the initial density matrix given in eqn (\ref{initialrhouncoupled}). However, it turns out that if one considers the averages of the initial values with respect to the following shifted (coupled) initial equilibrium density matrix of the bath in canonical ensemble, 
\begin{equation}\label{initialrhocoupled}
\hat{\rho}_B^{eq}=\frac{1}{Z}\exp\Big\lbrack -\beta \sum_j \Big\lbrace\frac{p_j^{2}(0)}{2m_j}+\frac{1}{2}m_j\omega_j^{2}\Big(q_j(0)-\frac{c_j x(0)}{m_j\omega_j^2}\Big)^2\Big\rbrace\Big\rbrack , 
\end{equation}
 then the operator \(F(t)\) gets associated with a Gaussian nature and one gets \(\langle F(t) \rangle = 0\). Typically, the initial preparation of the total system-plus-bath model fixes the statistical properties of the bath operators and the system's degrees of freedom. This initial preparation is the one which renders the fluctuating force \(F(t)\) with a Gaussian nature with time-homogenous correlations. Thus, the IST can be consistently absorbed into the quantum noise term if the initial density matrix of the bath is given by eqn (\ref{initialrhocoupled}). In other words, the appearance of the IST in the quantum Langevin equation is a consequence of the fact that in many treatments, the system and the bath are initially taken to be uncoupled, i.e. the initial equilibrium density matrix of the bath is given by eqn (\ref{initialrhouncoupled}). If however, the system and the bath are coupled from the initial instant, such a spurious term in the equation of motion can be effectively avoided \cite{bez,physicaA}.

\smallskip

Now, upon using linear response theory by considering a general c-number perturbing force $f(t)$ in addition to the random force, one can solve eqn (\ref{eqnm}) using a Fourier transform to give \cite{ford1,ford2},
\begin{equation}
\tilde{x}(\omega)=\alpha^{(0)}(\omega)[\tilde{f}(\omega)+\tilde{F}(\omega)]
\end{equation}
where,
\begin{equation}\label{alpha}
  \alpha^{(0)}(\omega) = \frac{1}{K-m\omega^2-i\omega\tilde{\mu}(\omega)}
\end{equation} is the generalized susceptibility and the Fourier transform of a dynamical variable is denoted by a tilde. The generalized susceptibility determines the dynamics of such a system in a unique way.

\smallskip

In order to characterize the details of a particular dissipation mechanism, one specifies the bath spectral function \(J(\omega)\) defined as,
\begin{equation}
  J(\omega) = \frac{\pi}{2} \sum_{j=1}^{N} \frac{c_j^2}{m_j \omega_j} \delta(\omega - \omega_j)
\end{equation} which then is equivalent to,
\begin{equation}
  \mu(t) = \frac{2}{\pi m} \int_{0}^{\infty} \frac{J(\omega)}{\omega} \cos (\omega t) d\omega.
\end{equation}
Thus, specifying the spectral properties of the bath degrees of freedom, i.e. specifying \(J(\omega)\) specifies the dissipation mechanism and consequently \(\alpha^{(0)}(\omega)\) (through \(\tilde{\mu}(\omega)\)). In this work, we consider three dissipation mechanisms namely, Ohmic dissipation, Drude dissipation and radiation bath dissipation.

\smallskip

If \(T\) is the temperature of the bath, then quantum thermodynamic properties of the dissipative quantum oscillator in the steady state is determined by the remarkable relation \cite{qtf1,qtf2},
\begin{equation}\label{Fmean}
  F(T) = \frac{1}{\pi} \int_{0}^{\infty} f(\omega,T) {\rm Im} \bigg[ \frac{d}{d \omega} \ln [\alpha^{(0)}(\omega + i0^+)] \bigg] d\omega
\end{equation} where \(F(T)\) denotes the free energy of the system at temperature \(T\) whereas, \(f(\omega,T)\) denotes the free energy of an oscillator of frequency \(\omega\) of the heat bath,
\begin{equation}\label{f}
  f(\omega,T) = k_B T \ln \Big[2\sinh\Big(\frac{\hbar\omega}{2k_BT}\Big)\Big].
\end{equation}
Note that the heat bath is in a Gibbs canonical state at temperature \(T\). Although how it reaches equilibrium at a given temperature is unimportant for us, we shall in all our subsequent discussions assume that this is indeed the case. Subsequently, eqn (\ref{f}) is the free energy of a single bath oscillator with the average taken over the canonical state of the bath. 

\smallskip

It appears from eqn (\ref{Fmean}) that as if in the steady state, the free energy of the system receives contributions from the bath oscillators whose frequencies range from $0$ to \(\infty\). The contribution coming from a particular frequency \(\omega\) is suitably weighed by the imaginary part of the derivative of the generalized susceptibility. By standard thermodynamic definitions, the mean energy of the system at temperature \(T\) is given by,
\begin{equation}
  E(T) = \frac{1}{\pi} \int_{0}^{\infty} \mathcal{E}(\omega,T) {\rm Im} \bigg[ \frac{d}{d \omega} \ln [\alpha^{(0)}(\omega + i0^+)] \bigg] d\omega
\end{equation} where \(\mathcal{E}(\omega,T)\) is the mean energy of a single bath oscillator with frequency \(\omega\). This greatly resembles the recent result (see for example \cite{jarzy4}) on the quantum analogue of energy equipartition for open quantum systems. According to this result, the mean energy of the system receives contributions from the bath degrees of freedom according to the relationship,
\begin{equation}
  E(T) = \int_{0}^{\infty}\mathcal{E}(\omega,T) \mathcal{P}(\omega) d\omega
\end{equation} where \(\mathcal{P}(\omega)\) is an appropriate probability distribution function, i.e. it is both positive definite and also normalized. With this in mind, we state the central result of our paper,
\begin{prop}\label{th1}
For the one dimensional dissipative quantum oscillator, the free energy is expressible in the form,
\begin{equation}
  F(T) := \langle f(\omega,T)\rangle = \int_{0}^{\infty} f(\omega,T) \mathcal{P}(\omega) d\omega.
\end{equation} where \(\mathcal{P}(\omega)\) is an appropriate probability distribution function. This means, one must identify,
  \begin{equation}\label{P}
    \mathcal{P}(\omega) = \frac{1}{\pi} {\rm Im} \bigg[ \frac{d}{d \omega} \ln [\alpha^{(0)}(\omega + i0^+)] \bigg]
  \end{equation} as a probability distribution function according to which the system receives contributions to its free energy from the degrees of freedom of the bath.
\end{prop}
Thus, one needs to show that $\mathcal{P}(\omega)$ must satisfy the following basic conditions,
\begin{enumerate}
\item Positivity:
\begin{equation}
\mathcal{P}(\omega) \geq  0 \hspace{5mm} \forall \hspace{2mm} \omega \in [0,\infty).
\end{equation}
\item Normalization:
\begin{equation}
\int_{0}^{\infty} \mathcal{P}(\omega) d\omega  =  1.
\end{equation}
\end{enumerate}

\smallskip

 Note that the net averaging is a two fold process. First one considers \(f(\omega,T)\) which is the average free energy of a single bath oscillator with the averaging taking place over the Gibbs canonical state of the heat bath. The second step is to sum over all the contributions to the mean free energy of the system coming from bath oscillators over the entire bath spectrum. The latter is performed with respect to the probability distribution function \(\mathcal{P}(\omega)\) related to the susceptibility. In the next section, we demonstrate that proposition-(\ref{th1}) is indeed true by considering different dissipation mechanisms. The result known in the literature as the quantum energy partition theorem follows from this as a straightforward corollary,
\begin{corollary}
  The mean energy of the dissipative quantum oscillator can be written in the following form,
  \begin{equation}\label{Eavg}
    E(T) = \int_{0}^{\infty} \mathcal{E}(\omega,T) \mathcal{P}(\omega)d\omega
  \end{equation} where \(\mathcal{E}(\omega,T)\) is the mean kinetic energy of an individual bath oscillator and \(\mathcal{P}(\omega)\) is a probability distribution function [eqn (\ref{P})] according to which the bath degrees of freedom subscribe to the mean energy of the system.
\end{corollary}
 For obvious reasons, we shall call the result present as proposition-(\ref{th1}), free energy partition for open quantum systems. This leads to a similar statement for the entropy of the dissipative oscillator,
 \begin{corollary}
  The entropy of the one dimensional dissipative quantum oscillator at temperature \(T\) is given by,
  \begin{equation}
    S(T) = \int_{0}^{\infty} s(\omega,T) \mathcal{P}(\omega) d\omega
  \end{equation} where \(s(\omega,T) = -\frac{\partial f(\omega,T)}{\partial T}\) is the entropy of a bath oscillator with frequency \(\omega\).
\end{corollary}
Noting that \(s(\omega,T) \rightarrow 0\) as \(T \rightarrow 0\) means \(\lim_{T \rightarrow 0}  S(T) = 0\) consistently reproducing the third law of thermodynamics in the quantum regime \cite{mkb2}.

\section{Free energy partition for different dissipation mechanisms}\label{section3}
In this section, we demonstrate that proposition-(\ref{th1}) is correct by considering three different examples of dissipation mechanisms. Let us first recall that the generalized susceptibility \(\alpha(\omega)\) has poles at its normal mode frequencies \(\{\tilde{\omega}_k\}\) and zeroes at the bath frequencies \(\{\omega_j\}\). Therefore, one can express the susceptibility as (see \cite{qtf3} and references therein),
\begin{equation}\label{normalmodealpha}
  \alpha^{(0)}(\omega) = - \frac{1}{m} \frac{\Pi_{j=1}^{N} (\omega^2 - \omega_j^2)}{\Pi_{k=0}^{N} (\omega^2 - \tilde{\omega}_k^2)}
\end{equation}
where the denominator is the product over normal modes of the Brownian oscillator and
the numerator is the product over those of the bath oscillators. One must notice that the number of distinct normal mode frequencies of the system is one more than that of the number of distinct bath frequencies. The normal mode frequencies are all real \cite{qtf3}. Furthermore, the zeros and poles of $\alpha(\omega)$ are all simple. We use these facts to prove the positivity of the quantum probability distribution function associated with the free energy of the Brownian oscillator. Let us prove the nonnegativity associated with the quantum probability distribution function $P(\omega)$ as mentioned in proposition-(\ref{th1}). Plugging the form of susceptibility as given in eqn (\ref{normalmodealpha}) in the quantum probability distribution function associated with the free energy [as given in eqn (\ref{P})], we obtain,
\begin{eqnarray}\label{positivityP}
\mathcal{P}(\omega)&=&\frac{1}{\pi}{\rm Im} \Bigg[ \bigg( \frac{1}{\sum_{j=1}^{N}(\omega-{\omega}_j)}+\frac{1}{\sum_{j=1}^{N}(\omega+{\omega}_j)}\bigg) \nonumber
 \\ &-&\bigg( \frac{1}{\sum_{k=0}^{N} (\omega- \tilde{\omega}_k)}+ \frac{1}{\sum_{k=0}^{N}(\omega+\tilde{\omega}_k)}\bigg) \Bigg].
\end{eqnarray}
Now, utilizing the identity,
\begin{equation}\label{deltafunctionidentity}
\frac{1}{x+i0^+}=P\Big(\frac{1}{x}\Big)-i\pi\delta(x)
\end{equation}
one can easily show that,
\begin{equation}\label{positiveP}
\mathcal{P}(\omega)=\sum_{k=0}^{N}[\delta(\omega-\tilde{\omega}_k)+\delta(\omega+\tilde{\omega}_k)]-\sum_{j=1}^{N}[\delta(\omega-{\omega}_j)+\delta(\omega+{\omega}_j)].
\end{equation}
This is always a positive quantity. In the next few subsections, we explicitly demonstrate the normalization of \(\mathcal{P}(\omega)\) for the Brownian oscillator for different bath spectrums. Let us begin with the simplest example, namely, Ohmic dissipation.

\subsection{Ohmic dissipation}
In case of Ohmic dissipation, the friction carries no memory at all, i.e. \(\mu(t) = m \gamma_0
\delta(t)\) where \(\gamma_0\) is the damping constant related to the mean relaxation time as \(\tau_\nu = 1/\gamma_0\). Consequently, one has \(\tilde{\mu}(\omega) = m \gamma_0\) and the bath spectral function reads, \(J(\omega) = m \gamma_0 \omega\). It should be remarked that this dissipation mechanism is UV divergent, i.e. some integrals involving \(\tilde{\mu}(\omega)\) or \(J(\omega)\) over the entire spectrum of bath frequencies diverge if suitable regularization is not performed. In particular, the mean energy of the dissipative oscillator diverges in this case (see for example \cite{ohmicdiverge,jarzy5}). For this reason, the significance of strictly Ohmic dissipation (without any regularization) is often questioned. However, it serves as an easy to analyze toy model for a dissipation mechanism whose characteristics resemble classical Brownian motion. We shall therefore first look at Ohmic dissipation owing to its sheer simplicity before moving on to more realistic dissipation mechanisms in subsequent subsections.

\smallskip

Now, for Ohmic dissipation, one has,
\begin{equation}
  \alpha^{(0)}(\omega) = \frac{-1}{m(\omega^2 - \omega_0^2 + i \gamma_0 \omega)}
\end{equation} where \(K = m \omega_0^2\) and this gives the function \(\mathcal{P}(\omega)\) to be,
\begin{equation}
\mathcal{P}(\omega)=\frac{1}{\pi} {\rm Im} \bigg[ \frac{d}{d \omega} \ln [\alpha^{(0)}(\omega)] \bigg] = \frac{\gamma_0}{\pi} \frac{(\omega^2 + \omega_0^2)}{(\omega^2 - \omega_0^2)^2 + \omega^2 \gamma_0^2}.
\end{equation} This is clearly positive definite consistent with the fact that we have asserted that \(\mathcal{P}(\omega)\) represents a probability density function. However, in order to qualify as a genuine probability distribution function, \(\mathcal{P}(\omega)\) must in addition be normalized. To show that, we consider re-expressing \(\mathcal{P}(\omega)\) as,
\begin{equation}
  \mathcal{P}(\omega) = \frac{1}{\pi} \bigg( \frac{z_+}{\omega^2 + z_+^2} + \frac{z_-}{\omega^2 + z_-^2}\bigg)
\end{equation} where,
\begin{equation}
  z_{\pm} = \frac{\gamma_0}{2} \pm \sqrt{\omega_0^2 - i\frac{\gamma_0^2}{4}} .
\end{equation}
Then, it is easy to show that,
\begin{equation}
   \int_{0}^{\infty} \bigg( \frac{z_+}{\omega^2 + z_+^2} + \frac{z_-}{\omega^2 + z_-^2}\bigg) d\omega = \bigg( \frac{\pi}{2} + \frac{\pi}{2}\bigg) = \pi
\end{equation} thus giving,
\begin{equation}
  \int_{0}^{\infty} \mathcal{P}(\omega) d\omega = 1.
\end{equation}
The fact that the function \(\mathcal{P}(\omega)\) is both positive definite and normalized allows it to be interpreted as a probability distribution function which weighs the contributions to the system's free energy from the bath degrees of freedom. In other words, \(f(\omega,T)\mathcal{P}(\omega)d\omega\) corresponds to the free energy contribution to the dissipative quantum oscillator's free energy coming from the bath oscillators lying within the frequency range \(\omega\) to \(\omega + d\omega\).

\subsection{Drude dissipation}
After having considered Ohmic dissipation as an easy to solve toy model, let us consider a more realistic dissipation mechanism. The Drude dissipation mechanism is characterized by,
\begin{equation}
  \mu(t) = m \Omega \gamma_0 e^{-\Omega t}
\end{equation} or equivalently,
\begin{equation}
  \tilde{\mu}(\omega) = \frac{m \gamma_0 \Omega}{\Omega - i \omega}
\end{equation} where \(\Omega\) is a cut-off frequency scale. It is a single relaxation time dissipation mechanism where \(\tau_c = 1/\Omega\) plays the role of a characteristic memory time scale over which the friction kernel decays. Introducing a new set of parameters \(\{\Gamma,\Omega_0,\Omega'\}\) related to the original parameters \(\{\gamma_0,K,\Omega\}\) through the following relations \cite{qtf2},
\begin{equation}\label{parameterDrude}
 \gamma_0 = \frac{m \Gamma[\Omega' (\Omega' + \Gamma) + \Omega_0^2]}{(\Omega' + \Gamma)^2}, \hspace{2mm} \frac{K}{m} = \frac{\Omega_0^2 \Omega'}{\Omega' + \Gamma}, \hspace{2mm} \Omega = \Omega' + \Gamma
\end{equation} the generalized susceptibility can be re-expressed as,
\begin{equation}
  \alpha^{(0)}(\omega) = - \frac{1}{m} \frac{\omega + i (\Omega' + \Gamma)}{(\omega + i \Omega')(\omega^2 + i \Gamma \omega - \Omega_0^2)}.
\end{equation}
This gives two equivalent expressions for \(\mathcal{P}(\omega)\),
\begin{equation}\label{PDrude1}
  \mathcal{P}(\omega) =\frac{1}{\pi} \Bigg[ \frac{-(\Omega' + \Gamma)}{\omega^2 + (\Omega' + \Gamma)^2} + \frac{\Omega'}{\omega^2 +\Omega'^2} + \frac{\Gamma (\omega^2 + \Omega_0^2)}{(\omega^2 - \Omega_0^2)^2 + \Gamma^2 \omega^2} \Bigg]
\end{equation}
and,
\begin{equation}\label{PDrude2}
  \mathcal{P}(\omega) = \frac{1}{\pi} \Bigg[ \frac{\Omega'}{\omega^2 + \Omega'^2} +  \frac{z_+}{\omega^2 + z_+^2}  +  \frac{z_-}{\omega^2 + z_-^2} - \frac{\Omega}{\omega^2 + \Omega^2} \Bigg]
\end{equation} where the latter has been expressed in terms of the frequencies,
\begin{equation}\label{normalmode}
  z_{\pm} = \frac{\Gamma}{2} \pm \sqrt{\Omega_0^2 - \frac{\Gamma^2}{4}}i.
\end{equation} Using eqn (\ref{PDrude2}), it is straightforward to check that,
\begin{equation}
  \int_{0}^{\infty} \mathcal{P}(\omega)d\omega = \frac{1}{\pi} \bigg[ \frac{\pi}{2} + \frac{\pi}{2} + \frac{\pi}{2} - \frac{\pi}{2} \bigg] = 1.
\end{equation}
Therefore \(\mathcal{P}(\omega)\) given in eqns (\ref{PDrude1}) or (\ref{PDrude2}) represents the appropriate probability distribution function according to which free energy partition takes place in the dissipative quantum oscillator.

\subsection{Radiation bath}
Our final example is that of the radiation bath. This kind of dissipation mechanism natural arises when the bath is composed of a black body radiation field \cite{qtf1}. For this particular case, the dissipation kernel reads,
\begin{equation}
  \mu(t) = \frac{2e^2 \Omega^2}{3 c^3} [ 2\delta(t) - \Omega e^{-\Omega t}]
\end{equation} or equivalently,
\begin{equation}
  \tilde{\mu}(\omega) = \frac{2e^2 \omega \Omega^2}{3 c^3 (\omega + i \Omega)} = \frac{M \omega \Omega}{(\omega + i \Omega)}
 \end{equation} where \(\Omega\) plays the role of a cut-off frequency scale characteristic to the bath and \(M\) is a renormalized mass given by \(M = m + \frac{2 e^2 \Omega}{3 c^3} \approx \frac{2 e^2 \Omega}{3 c^3}\) in the large cut-off limit \((\Omega \rightarrow \infty)\). Upon defining the parameters \(\{\Gamma,\Omega',\Omega_0\}\) using relations \cite{qtf2,PRL55},
 \begin{equation}\label{parameterRadiation}
   \frac{1}{\Omega} = \frac{1}{\Omega'} + \frac{\Gamma}{\Omega_0^2}, \hspace{2mm} \frac{K}{M} = \frac{\Omega_0^2 \Omega'}{\Omega' + \Gamma}, \hspace{2mm} \frac{M}{m} = \frac{(\Omega_0^2 + \Gamma \Omega')(\Omega' + \Gamma)}{\Omega_0^2 \Omega'}
 \end{equation} one finds the following expression for the generalized susceptibility,
 \begin{equation}
   \alpha^{(0)}(\omega) = -\frac{1}{m} \frac{\omega + i \Omega}{(\omega + i \Omega')(\omega^2 + i \Gamma \omega - \Omega_0^2)}.
 \end{equation}
 This immediately gives the following two equivalent forms in which the function \(\mathcal{P}(\omega)\) can be expressed,
 \begin{equation}\label{Pradiation1}
   \mathcal{P}(\omega) = \frac{1}{\pi} \Bigg[ -\frac{\Omega}{\omega^2 + \Omega^2} + \frac{\Omega'}{\omega^2 + {\Omega'}^2} + \frac{\Gamma(\omega^2 + \Omega_0^2)}{(\omega^2 - \Omega_0^2)^2 + \Gamma^2 \omega^2}\Bigg]
 \end{equation} and,
 \begin{equation}\label{Pradiation2}
   \mathcal{P}(\omega) = \frac{1}{\pi} \Bigg[ -\frac{\Omega}{\omega^2 + \Omega^2} + \frac{\Omega'}{\omega^2 + {\Omega'}^2} + \frac{z_+}{\omega^2 + z_+^2} + \frac{z_-}{\omega^2 + z_-^2}\Bigg]
 \end{equation} where \(z_{\pm}\) are defined using the same relations as eqn (\ref{normalmode}). Normalization of \(\mathcal{P}(\omega)\) can be checked by integrating eqn (\ref{Pradiation2}) giving,
\begin{equation}
  \int_{0}^{\infty} \mathcal{P}(\omega) d\omega = \frac{1}{\pi} \Bigg[ -\frac{\pi}{2} + \frac{\pi}{2} + \frac{\pi}{2} + \frac{\pi}{2} \Bigg] = 1.
\end{equation} Thus, in summary, the function \(\mathcal{P}(\omega)\) defined from the generalized susceptibility using eqn (\ref{P}) indeed fulfills the basic requirements needed to qualify as a probability distribution function, in accordance with proposition-(\ref{th1}).

\smallskip

\begin{figure}
\begin{center}
\includegraphics[scale=0.80]{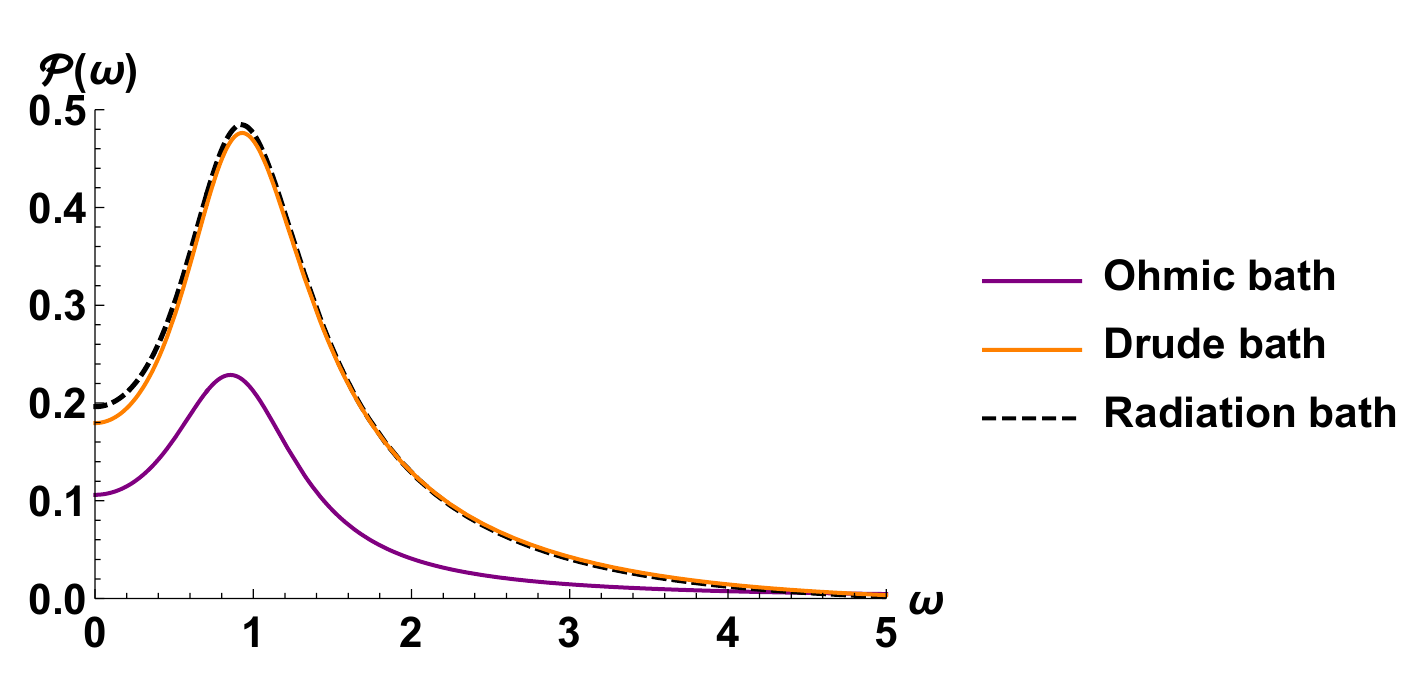}
\caption{Plot of the probability distribution function \(\mathcal{P}(\omega)\) as a function of \(\omega\) for: (a) Ohmic dissipation (violet) with \(\Gamma = 3\) and \(\Omega_0 = 2\); (b) Drude dissipation (orange) with \(\Gamma = 3\), \(\Omega_0 = 2\) and \(\Omega = 5\); (c) Radiation bath (dashed) with \(\Gamma = 3\), \(\Omega_0 = 2\) and \(\Omega = 5\).}
\label{one}
\end{center}
\end{figure}

The probability distribution function \(\mathcal{P}(\omega)\) is plotted as a function of \(\omega\) for the three different dissipation mechanisms in figure-(\ref{one}). A common theme is the dominance of contributions due to the low frequency modes of the bath. In general, there is a most probable frequency \(\omega = \omega_m\) from which there is maximum contribution to the system's free energy. Contributions from the high frequency modes are significantly less.

\smallskip

 For the radiation bath, a particularly interesting case is the large cut-off limit \(\Omega \rightarrow \infty\) wherein one gets \cite{PRL55},
 \begin{equation}
  	\frac{1}{\pi} {\rm Im} \bigg[\frac{d}{d\omega} \ln [\alpha_0(\omega)] \bigg] = \frac{1}{\pi} \Bigg[\frac{\Gamma(\omega^2 + \Omega_0^2)}{(\omega^2 - \Omega_0^2)^2 + \Gamma^2 \omega^2} - \frac{\Gamma}{\Omega_0^2}\Bigg].
\end{equation}
Here we identify as \(\mathcal{P}(\omega)\), the first term, i.e.
\begin{equation}
\mathcal{P}(\omega) = \frac{1}{\pi}\frac{\Gamma(\omega^2 + \Omega_0^2)}{(\omega^2 - \Omega_0^2)^2 + \Gamma^2 \omega^2}
\end{equation} which is both positive definite and can be easily checked to be normalized. The second term (\(-\Gamma/\Omega_0^2\)) leads to a quantum electrodynamic correction to the oscillator's free energy \cite{PRL55}.

\section{Generalization to three dimensional systems}\label{section4}
\begin{figure}
\begin{center}
\includegraphics[scale=0.80]{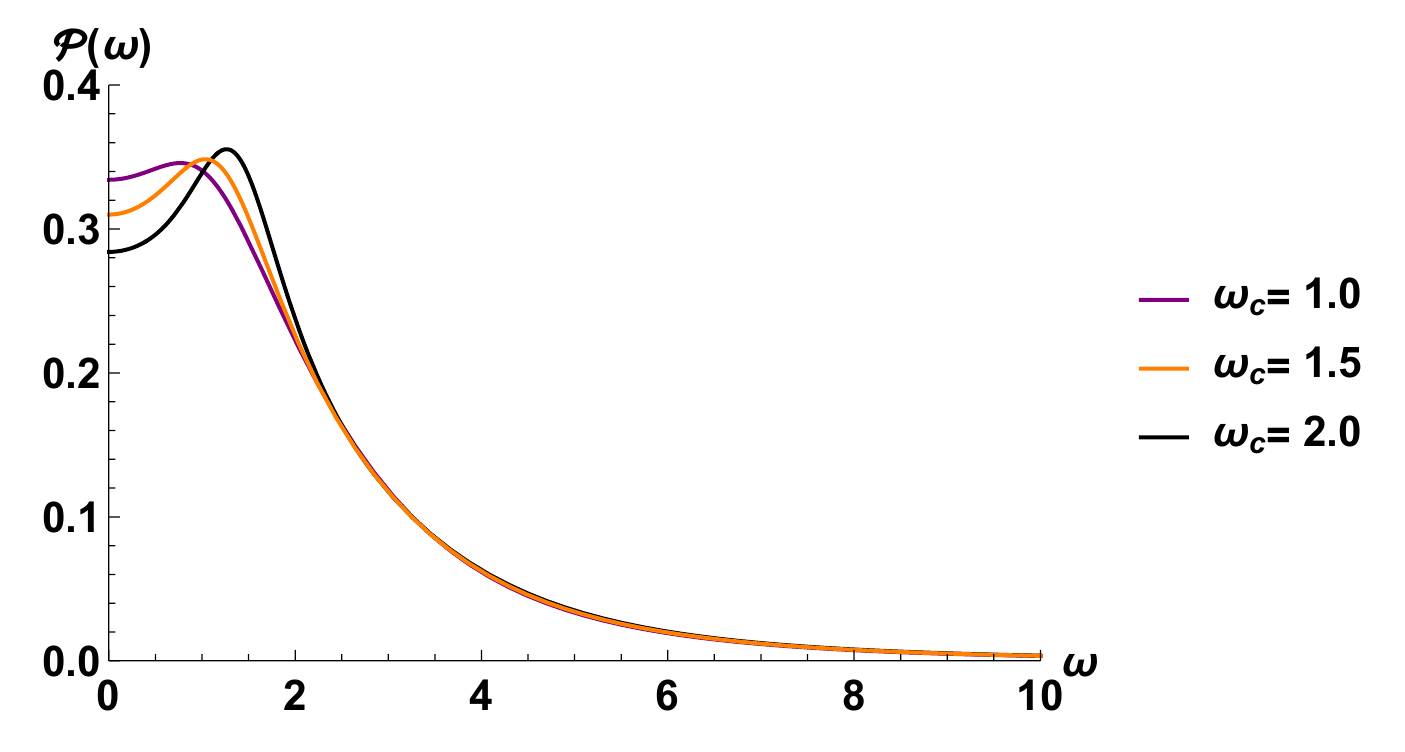}
\caption{Plot of \(\mathcal{P}(\omega)\) as a function of \(\omega\) for Drude dissipation with \(\Gamma = 3\), \(\Omega_0 = 2\) and \(\Omega =5\) for different values of magnetic field \(\omega_c\).}
\label{two}
\end{center}
\end{figure}

The novel formalism proposed in section-(\ref{section2}) where extensive thermodynamic quantities of the dissipative system are obtained as a two fold average of similar quantities associated with the bath degrees of freedom can be generalized to real world three dimensional systems. The probability distribution \(\mathcal{P}(\omega)\) is associated with the generalized susceptibility of the system, a quantity which can be measured in experiments by applying small perturbations to the dissipative system. Thus, a generalization of our results to three dimensions brings the theoretical formalism closer to experimental verification possibly opening new directions. In the absence of external fields, all the three directions of motion are independent so that the free energy of the dissipative oscillator is simply eqn (\ref{Fmean}) with \(f(\omega,T)= 3 k_B T \ln [2 \sinh(\frac{\hbar\omega}{2k_BT})] \) being the mean free energy of a three dimensional bath oscillator. However, in a more general case where one imposes an external magnetic field, the mean free energy of the system acquires additional contributions due to this external field. The free energy in this case is given by \cite{ford3,ford4},
\begin{eqnarray}
  F(T,B) &=&\frac{1}{\pi}\int_{0}^{\infty}d\omega f(\omega,T){\rm Im}\Big[\frac{d}{d\omega}\ln[\det\alpha(\omega+i0^+)]\Big] \\
  &=& F_0(T) + \Delta F(T,B)
\end{eqnarray} where the first term is simply three times of eqn (\ref{Fmean}) while the second term carries the magnetic field contribution and is given by,

\begin{equation}\label{DeltaF}
  \Delta F(T,B) = - \frac{1}{3\pi} \int_{0}^{\infty} f(\omega,T) {\rm Im} \Bigg[ \frac{d}{d\omega} \ln \bigg(1 - \bigg(\frac{eB \omega \alpha^{(0)}(\omega)}{c}\bigg)^2 \bigg) \Bigg] d\omega.
\end{equation}
 Here, \(\alpha^{(0)}(\omega)\) is the susceptibility tensor defined in eqn (\ref{alpha}). Let us note that in this case, the bath oscillators are three dimensional, which can be seen right from the associated Hamiltonian,
\begin{eqnarray}
   H &=& \frac{(\mathbf{p} - \frac{e \mathbf{A}}{c})^2}{2m} + \frac{1}{2} K \mathbf{r}^2 \nonumber \\
   && + \sum_{j}\bigg[\frac{\mathbf{p}_j^2}{2m_j} + \frac{1}{2}m_j \omega_j^2 \bigg( \mathbf{q}_j - \frac{c_j}{m_j \omega_j^2}\mathbf{r} \bigg)^2 \bigg]
\end{eqnarray}
which gives the correct three dimensional quantum Langevin equation when the bath variables are eliminated \cite{12}. Consequently, \(f(\omega,T)\) is three times the expression given in eqn (\ref{f}). This results in the factor of \(1/3\) in eqn (\ref{DeltaF}). For all the three dissipation mechanisms considered in the previous section, \(\alpha(\omega)\) can be expressed as \cite{ford3,ford4},
\begin{equation}
\alpha(\omega)=\frac{\omega+i\Omega}{m(\omega+i\Omega)(\Omega_0^2-\omega^2-i\Gamma\omega)}.
\end{equation}
where the Ohmic bath is recovered under the limits: \(\Omega_0 \rightarrow \omega_0\), \(\Omega \rightarrow \infty\) and \(\Gamma \rightarrow \gamma_0/m\). The parameters \(\{\Omega,\Omega_0,\Gamma\}\) for Drude and radiation baths are the ones defined by eqns (\ref{parameterDrude}) and (\ref{parameterRadiation}) respectively. Using this, one gets the following expression for the free energy,

\begin{eqnarray}
%\hskip-1.0cm
F(T,B)&=&\frac{3k_BT}{\pi}\int_0^{\infty}d\omega\ln\Big[2\sinh\Big(\frac{\hbar\omega}{2k_BT}\Big)\Big]\Big(-\frac{\Omega}{\omega^2+\Omega^2}+\frac{\Omega'}{\omega^2+{\Omega'}^2}\Big) \nonumber \\
%\hskip-1.0cm
&+&\frac{k_BT}{\pi}\int_0^{\infty}d\omega\ln\Big[2\sinh\Big(\frac{\hbar\omega}{2k_BT}\Big)\Big]\Big(\frac{z_+}{\omega^2+ {z_+}^2}+\frac{{z_-}}{\omega^2+{z_-}^2}+\frac{\Omega_1}{\omega^2+\Omega_1^2} \nonumber \\
&+&\frac{\Omega_1^*}{\omega^2+{\Omega_1^*}^2} +\frac{\Omega_2}{\omega^2+{\Omega_2}^2}+\frac{\Omega_2^*}{\omega^2+{\Omega_2^*}^2}\Big) \label{F3dim}
\end{eqnarray}

where \(z_{\pm}\) are defined using eqns (\ref{normalmode}) and,
\begin{eqnarray}
  \Omega_1 &=& \Big\lbrack\frac{\Gamma}{2}+\Big(\frac{b-a}{2}\Big)^{\frac{1}{2}}\Big\rbrack-i\Big\lbrack\frac{\omega_c}{2}+\Big(\frac{b+a}{2}\Big)^{\frac{1}{2}}\Big\rbrack, \nonumber \\
  \Omega_2&=&\Big\lbrack\frac{\Gamma}{2}-\Big(\frac{b-a}{2}\Big)^{\frac{1}{2}}\Big\rbrack-i\Big\lbrack\frac{\omega_c}{2}-\Big(\frac{b+a}{2}\Big)^{\frac{1}{2}}\Big\rbrack
\end{eqnarray}
  with,
  \begin{equation}
    a =\Big(\frac{\omega_c}{2}\Big)^2+\Big(\Omega_0^2-\frac{\Gamma^2}{4}\Big) \hspace{3mm} {\rm and}  \hspace{3mm} b=\Big\lbrack a^2+\Big(\frac{\Gamma\omega_c}{2}\Big)^2\Big\rbrack^{\frac{1}{2}}.
  \end{equation}
  Upon identifying \(f(\omega,T) = 3 k_B T \ln [2 \sinh(\frac{\hbar\omega}{2k_BT})\)] as the free energy of an individual three dimensional bath oscillator, one finds the following expression for \(\mathcal{P}(\omega)\) so that eqn (\ref{F3dim}) can be expressed in the form of eqn (\ref{Fmean}),

\begin{eqnarray}
\mathcal{P}(\omega) = \frac{1}{\pi} \Bigg[-\frac{\Omega}{\omega^2+\Omega^2} &+& \frac{\Omega'}{\omega^2+{\Omega'}^2} + \frac{1}{3}\bigg( \frac{z_+}{\omega^2+ {z_+}^2}+\frac{{z_-}}{\omega^2+{z_-}^2} + \frac{\Omega_1}{\omega^2+\Omega_1^2} \nonumber \\
&+&\frac{\Omega_1^*}{\omega^2+{\Omega_1^*}^2}+\frac{\Omega_2}{\omega^2+{\Omega_2}^2}+\frac{\Omega_2^*}{\omega^2+{\Omega_2^*}^2}\bigg)\Bigg] \label{Pthreedim}.
\end{eqnarray}
Now integrating \(\mathcal{P}(\omega)\) gives,
\begin{equation}
  \int_{0}^{\infty} \mathcal{P}(\omega) d\omega = \frac{1}{\pi} \Bigg[ -\frac{\pi}{2} + \frac{\pi}{2} + \frac{1}{3} \bigg( \frac{\pi}{2} + \frac{\pi}{2} + \frac{\pi}{2}+ \frac{\pi}{2} + \frac{\pi}{2} + \frac{\pi}{2} \bigg) \Bigg] = 1
\end{equation} which means that \(\mathcal{P}(\omega)\) is exactly normalized. Henceforth, the function \(\mathcal{P}(\omega)\) characterizes the probabilities according to which the bath oscillators contribute to the mean free energy of the dissipative magneto-oscillator. One can therefore state a more general form of proposition-(\ref{th1}),
\begin{prop}
  For a three dimensional dissipative magneto-oscillator, the free energy can be expressed as the following two fold average,
  \begin{equation}
   F(T) := \langle f(\omega,T)\rangle = \int_{0}^{\infty} f(\omega,T) \mathcal{P}(\omega) d\omega
  \end{equation} where \(f(\omega,T)\) is the free energy of an individual three dimensional bath oscillator and \(\mathcal{P}(\omega)\) is a probability distribution function dependent on \(B\) as,
  \begin{eqnarray}\label{Pgenthreedim}
     \mathcal{P}(\omega) &=& \frac{1}{\pi} {\rm Im} \bigg[ \frac{d}{d \omega} \ln [\det \alpha(\omega+i0^+)] \bigg]
  \end{eqnarray}
\end{prop}
The positivity of $\mathcal{P}(\omega)$ can be demonstrated as follows. Following \cite{ford4}, one can show that,
\begin{equation}\label{detalpha}
  \det\alpha(\omega) \propto \frac{\Pi_{j=1}^{N}(\omega^2-\omega_j^2)}{\Pi_{k=0}^{N}(\omega^2-\tilde{\omega}_k^2)}
\end{equation}
where $\omega_j$ and $\tilde{\omega}_k$ follow the same nomenclature as that of eqn (\ref{normalmodealpha}). Again using the identity of eqn (\ref{deltafunctionidentity}), one can show that,
\begin{eqnarray}
\mathcal {P}(\omega) = \sum_{k=0}^{N}[\delta(\omega-\tilde{\omega}_k)] -\sum_{j=1}^{N}[\delta(\omega-\omega_j)].
\end{eqnarray}
Therefore with the same reasoning as before, one finds that $\mathcal {P}(\omega)$ is positive definite. Thus, the field dependent probability distribution function \(\mathcal{P}(\omega)\) has been verified to be positive definite and exactly normalized for Ohmic, Drude and radiation baths. Eqn (\ref{Pgenthreedim}) therefore appropriately generalizes proposition-(\ref{th1}) to three dimensions in the presence of an external magnetic field. It is easy to see that for \(B = 0\), eqn (\ref{Pgenthreedim}) coincides with eqn (\ref{P}) with \( f(\omega, T) \)  being the free energy of an individual three (rather than one) dimensional bath oscillator. For the three dimensional case, the probability distribution function \(\mathcal{P}(\omega)\) is plotted as a function of \(\omega\) for various external magnetic field strengths in figure-(\ref{two}). The case of Drude dissipation is taken for the purpose of illustration. An immediate observation on the effect of the magnetic field is an increase in the numerical value of the most probable frequency. Furthermore, the magnetic field suppresses contributions to the mean free energy of the system from very low energy bath modes. Upon increasing the magnetic field, the peak of \(\mathcal{P}(\omega)\) gets sharper and therefore, the contributions from frequencies away from the most probable value diminishes with an increase in magnetic field. Nevertheless, the bath modes with frequencies smaller than the most probable value are more significant than the modes with frequencies higher than the most probable value.

\section{Conclusions}\label{section5}
In this work, we have uncovered a novel aspect of quantum thermodynamics in dissipative systems. It has been demonstrated that after the system has attained equilibrium with respect to the bath, the system's mean free energy and hence, several other thermodynamic functions such as energy and entropy can be interpreted as being composed of contributions coming from the heat bath oscillators distributed over the entire frequency spectrum. This is mathematically expressed as \(F(T) = \langle f(\omega,T) \rangle\) where \(f(\omega,T)\) is the average free energy of a single bath oscillator with frequency \(\omega\) at temperature \(T\). Such an average is obtained over the Gibbs canonical state of the bath. The averaging denoted by \(\langle \cdot \rangle\) is performed at the second stage wherein contributions to \(F(T)\) due to bath oscillators with different frequencies are summed over a well defined probability distribution function \(\mathcal{P}(\omega)\) which is both positive definite and normalized. The general form of this function \(\mathcal{P}(\omega)\) has been proposed in the previous section in eqn (\ref{Pgenthreedim}) where the possibility of an external magnetic field is also taken into account. This result has been termed as free energy partition and it is this novel interpretation of quantum thermodynamic functions which naturally leads to both the quantum energy partition theorem and the third law of thermodynamics. As yet another corollary, one can express the heat capacity of such a dissipative system in the form,
\begin{equation}
  C(T) := \langle c (\omega,T) \rangle = \int_{0}^{\infty} c (\omega,T) \mathcal{P}(\omega) d\omega
\end{equation} where \(c (\omega,T) = -T \partial^2 f/\partial T^2 = \partial \mathcal{E}/\partial T\) is the heat capacity associated with a single bath oscillator. In the context of the specific heat however, one must keep in mind some subtle issues reported earlier in the literature \cite{hanggi} (see also \cite{jarzy31}). It has been demonstrated that for the free (quantum) Brownian particle, in the case of Drude dissipation, the specific heat can take negative values if the damping constant \(\gamma_0\) exceeds the Drude cut-off frequency \(\Omega\). However, this issue shall not arise if the system (the Brownian particle) is placed in a confining potential, i.e. one considers the dissipative oscillator instead of the free Brownian particle \cite{qtf4}. 

\smallskip

It should perhaps be emphasized that with the recent developments in experimental methods, one should be able to verify our results. In particular, we have related the probability distribution function \(\mathcal{P}(\omega)\) to the generalized susceptibility, which can be measured in experiments, by imposing small perturbations such as electric and magnetic fields. Since the same function dictates the distributions associated with multiple thermodynamic functions such as the free energy, entropy and internal energy, it must be of quite general significance. In particular, although an intimate connection of \(\mathcal{P}(\omega)\) with linear response theory and fluctuation-dissipation relations is evident, a more physically intuitive understanding of such functions needs to be understood in the context of general dissipative systems. As such, the present work should open up new possibilities and motivate further advances in the field of quantum thermodynamics.

\section*{Acknowledgements}
J.K. acknowledges the financial support received from IIT Bhubaneswar in the form of an Institute Research Fellowship. The work of A.G. is supported by the M.H.R.D., Government of India in the form of a Prime Minister's Research Fellowship. M.B. gratefully acknowledges financial support from the Department of Science and Technology (DST), India under the Core grant (Project No. CRG/2020//001768). Finally, we thank the anonymous referee for their valuable comments which have improved the paper.


\begin{thebibliography}{99}

\bibitem{qt1}J. Gemmer, M. Michel and G. Mahler, {\it Quantum Thermodynamics},
Springer (2009).

\bibitem{qt2}S. Vinjanampathy and J. Anders, {\it Quantum Thermodynamics},
Contemporary Physics, 57, 545 (2016).

\bibitem{qtf1}G. W. Ford, J. T. Lewis and R. F. O'Connell, {\it  Quantum oscillator in a blackbody radiation field
II. Direct calculation of the energy using the fluctuation-dissipation theorem},
Ann. Phys. 185, 270 (1988).

\bibitem{qtf2}G. W. Ford and R. F. O'Connell, {\it Quantum thermodynamic functions for an oscillator coupled to
a heat bath},  Phys. Rev. B 75, 134301 (2007).

\bibitem{qtf3}M. Bandyopadhyay, {\it Quantum thermodynamics of a charged magneto-oscillator coupled to a heat bath},
J. Stat. Mech. (2009) 05002.

\bibitem{qtf4} S. Dattagupta, J. Kumar, S. Sinha and P. A. Sreeram, {\it Dissipative quantum systems and the heat capacity},
Phys. Rev. E 81, 031136 (2010). 

\bibitem{tl1}A. Shastry, Y. Xu and C. A. Stafford, {\it The third law of thermodynamics in open quantum systems},
J. Chem. Phys. 151, 064115 (2019).


%%%%%%%%%%%%%%%%%%%%%%%%%%%%%%%%

\bibitem{jarzy0}P. Bialas and J. Łuczka, {\it Kinetic Energy of a Free Quantum Brownian Particle}, Entropy 20, 123 (2018).

\bibitem{jarzy1}J. Spiechowicz, P. Bialas and J. Łuczka, {\it Quantum partition of energy for a free Brownian particle: Impact of dissipation}, Phys. Rev. A 98, 052107 (2018).

\bibitem{jarzy2}P. Bialas, J. Spiechowicz and J. Łuczka, {\it Partition of energy for a dissipative
quantum oscillator}, Sci. Rep. 16080 (2018).

\bibitem{jarzy3}P. Bialas, J. Spiechowicz and J. Łuczka, {\it Quantum analogue of energy equipartition
theorem}, J. Phys. A: Math. Theor. 52, 15 (2019).

\bibitem{jarzy31} J. Spiechowicz and J. Łuczka, {\it On superstatistics of energy for a free quantum
Brownian particle}, J. Stat. Mech. 064002 (2019).

\bibitem{jarzy4}J. Łuczka, {\it Quantum Counterpart of Classical Equipartition of Energy}, J. Stat. Phys. 179, 839-845 (2020).

\bibitem{jarzy5} J. Spiechowicz and J. Łuczka, {\it Energy of a free Brownian particle coupled to thermal vacuum}, Sci Rep 11, 4088 (2021).

\bibitem{kaur}J. Kaur, A. Ghosh and M. Bandyopadhyay, {\it Quantum counterpart of energy equipartition theorem for a dissipative charged magneto-oscillator: Effect of dissipation, memory, and magnetic field}, Phys. Rev. E 104, 064112 (2021).

\bibitem{nano1}N. Taniguchi, {\it Quantum thermodynamics of nanoscale steady states far from equilibrium}, Phys. Rev. B 97, 155404 (2018).

\bibitem{nano2}G. Guarnieri, G. T. Landi, S. R. Clark and J. Goold, {\it Thermodynamics of precision in quantum nonequilibrium steady states}, Phys. Rev. Research 1, 033021 (2019).
    
\bibitem{nano3}Salil Bedkihal, Malay Bandyopadhyay, Dvira Segal, {\it Flux-dependent occupations and occupation difference in geometrically symmetric and energy degenerate double-dot Aharonov-Bohm interferometers}, Phys. Rev. B {\bf 87}, 045418 (2013)

\bibitem{12}G. W. Ford, J.T. Lewis and R.O O’Connell, {\it Quantum Langevin equation}, Phys. Rev.
A 37, 11 (1988).

\bibitem{12A}S. Gupta and M. Bandyopadhyay, {\it Quantum Langevin equation of a charged oscillator in a magnetic field and coupled to a heat bath through momentum variables},  Phys. Rev. E 84, 041133 (2011).
    


\bibitem{bez} W. Bez, {\it Microscopic preparation and macroscopic motion of a Brownian particle}, Z. Phys. B Condens. Matter 39 (1980) 319.

\bibitem{physicaA} J. S. Canizares and F. Sols, {\it Translational symmetry and microscopic preparation in oscillator models of quantum dissipation}, Physica A 212 (1994) 181-193.

\bibitem{ford1}X.L. Li and R.F. O'Connell, {\it Green's function and position correlation function for
a charged oscillator in a heat bath and a magnetic field}, Physica A 224, 639 (1996).

\bibitem{ford2}X. L. Li, G. W. Ford and R. F. O’Connell, {\it Dissipative effects on the localization of a charged oscillator in a magnetic field}, Phys. Rev. E 53, 3359 (1996).

\bibitem{mkb2}M. Bandyopadhyay, {\it Dissipative Cyclotron Motion of a Charged Quantum-Oscillator and Third Law}, J. Stat. Phys. 140, 603 (2010).

\bibitem{ohmicdiverge}H. Grabert, P. Schramm and G. -L. Ingold, {\it Quantum Brownian motion: The functional integral approach}, Phys. Rep. 168, 115
(1988).

\bibitem{PRL55} G. W. Ford, J.T. Lewis and R. F. O’Connell, {\it  Quantum Oscillator in a Blackbody Radiation Field}, Phys. Rev. Lett. 55, 2273 (1985).

\bibitem{ford3}X. L. Li, G. W. Ford and R. F. O’Connell, {\it  Magnetic-field effects on the motion of a charged
particle in a heat bath}, Phys. Rev. A 41, 5287 (1990).

\bibitem{ford4}X. L. Li, G. W. Ford and R. F. O’Connell, {\it  Charged oscillator in a heat bath in the presence of a
magnetic field}, Phys. Rev. A 42, 4519 (1990).

\bibitem{hanggi}P. Hanggi, G. -L. Ingold and P. Talkner, {\it  Finite quantum dissipation: the challenge
of obtaining specific heat}, New J. Phys. 10, 115008 (2008). 



\end{thebibliography}
\end{document}